\documentclass[aps,prl,twocolumn]{revtex4}
\usepackage{graphicx}
\usepackage{amsmath,amssymb,amsfonts}

\begin{document}

\title{Giant Intrinsic Spin-Orbit Coupling in Bilayer Graphene}

\author{Hai-Wen Liu$^{1}$}
\author{ X.C. Xie$^{1,2}$}
\author{ Qing-feng Sun$^{1}$}
\email{qfsun@aphy.iphy.ac.cn}

\affiliation {$^1$Institute of Physics, Chinese Academy of Sciences,
Beijing 100190,
China\\
$^2$Department of Physics, Oklahoma State University, Stillwater,
Oklahoma 74078}

\date{\today}

\begin{abstract}
The intrinsic spin-orbit coupling (ISOC) in bilayer graphene is
investigated. We find that the largest ISOC between $\pi$ electrons
origins from the following hopping processes: a $\pi$ electron hops
to $\sigma$ orbits of the other layer and further to $\pi$ orbits
with the opposite spin through the intra-atomic ISOC. The magnitude
of this ISOC is about $0.46meV$, 100 times larger than that of the
monolayer graphene. The Hamiltonians including this ISOC in both
momentum and real spaces are derived. Due to this ISOC, the
low-energy states around the Dirac point exhibit a special spin
polarization, in which the electron spins are oppositely polarized
in the upper and lower layers. This spin polarization state is
robust, protected by the time-reversal symmetry. In addition, we
provide a hybrid system to select a certain spin polarization state
by an electric manipulation.
\end{abstract}

\maketitle

Recently, ultrathin graphitic devices including monolayer and
bilayer graphene have been fabricated
experimentally\cite{ref1,ref2}. The experimental developments
greatly stimulate theoretical research on this subject\cite{ref3}.
Monolayer graphene is a gapless semiconductor with unique k-linear
and massless Dirac-like spectrum, while bilayer graphene presents
parabolic and massive spectrum. This distinction in dispersion leads
to many physical diversities. To name a few, the Hall plateaus in
monolayer graphene appear at half-integer values while that in
bilayer graphene appear at integer values without the zero-level
plateau\cite{ref1,ref2,ref4,ref5,ref6}. The monolayer graphene
exhibits the weak anti-localization behavior due to the $\pi$
Berry's phase, however, in bilayer graphene the $2\pi$ Berry's phase
leads to weak localization\cite{ref8}. The difference between
monolayer and bilayer graphene has led to great effort in
research.

The spin-orbit coupling (SOC) has attracted great attention in the
last decade as it plays a very important role in the field of
spintronics. SOC can couple the electron spin to its orbital motion
and vice versa, thereby giving a useful handle for manipulating the
electron spin by external electric fields. Furthermore, the SOC may
also generate some exciting phenomena, e.g. the spin Hall effect and
the topological insulator.\cite{she} The SOC origins from the
relativistic effect, so it is usually very weak. For example, in a
monolayer graphene the theoretical research both in tight-binding
model and the first principle calculation show that the strength of
SOC there is only of the order $10^{-6}eV$.\cite{ref12} It will be
very interesting to find a large SOC or to enhance the SOC in a
graphene system by some means.

\begin{figure}
\includegraphics[width=8cm,totalheight=6cm]{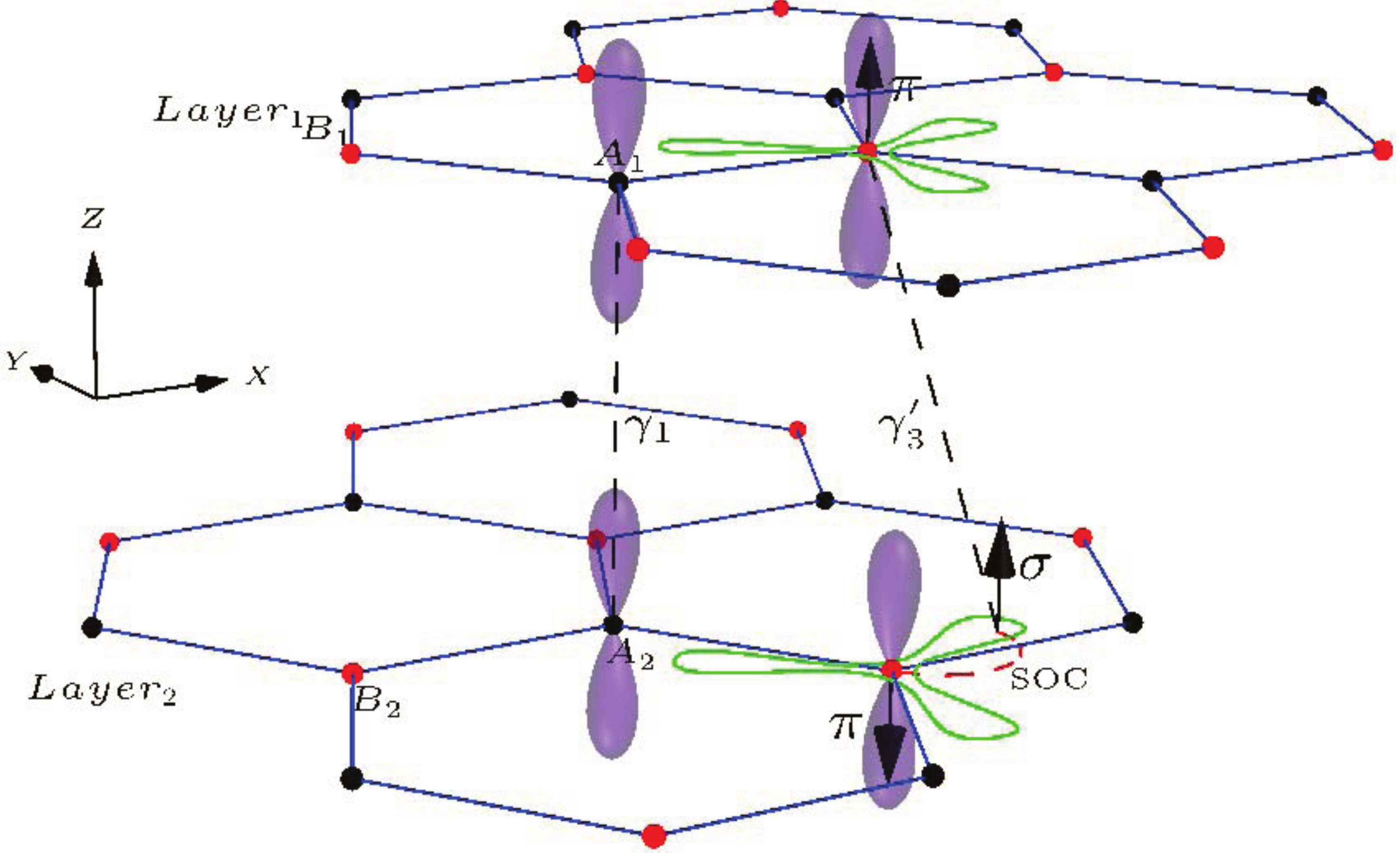}
\caption{(color online) Schematic diagram for hopping processes in
bilayer graphene, which leads the largest ISOC between $\pi$
electrons. The black (red) point stands for A (B) sublattice,
respectively.}
\end{figure}

In a carbon atom, the SOC $\Delta\vec{L}\cdot\vec{\sigma}$ (with the
angular momentum $\vec{L}$ and spin $\vec{\sigma}$) can couple the
$2p_z$ orbit (or $\pi$ orbit in the graphene) to the opposite-spin
$2p_{x}$ and $2p_{y}$ orbits (or $\sigma$ orbits in the graphene).
The magnitude $\Delta$ can be estimated to be of order $5meV$. In
monolayer graphene, $\sigma$ orbits hybridized in a $sp^{2}$
configuration to form the honeycomb lattice, while, the remaining
$2p_z$ orbit, which is perpendicular to the graphene plane, forms
the $\pi$ orbits and dominates the low energy physical properties of
graphene. Because the $\pi$ orbits are perpendicular to the $sp^{2}$
orbits, the hopping process from $\pi$ orbits to all $\sigma$ orbits
are suppressed. Therefore the intrinsic SOC (ISOC) in the low-energy
$\pi$ electrons must involve three hopping events: the SOC between
the spin-up $\pi$ orbit and the spin-down $\sigma$ orbit, the
hopping coupling from the $\sigma$ orbit to its nearest-neighbor
(N.N.) $\sigma$ orbit with the same spin, and the SOC from the
spin-down $\sigma$ orbit to the spin-up $\pi$ orbit in the N.N.
atom. This ISOC is on the order $\Delta^2$ and its strength is very
small, about $10^{-6}eV$,\cite{ref12}. There are several ways
proposed to enlarge the SOC strength,\cite{ref13,ref14} e.g. by
mixture of $\pi$ electrons with $\sigma$ electrons on the N.N. sites
by introducing impurities, but this approach will cause the system
to be unclean.

In this Letter, we predict a large ISOC in a clean bilayer graphene.
We consider bilayer graphene which consists of two coupled honeycomb
lattice with sublattice A1, B1 (A2, B2) on the top (bottom) layer,
respectively, arranged by Bernal stacking as illustrated in Fig. 1.
Every A1 site lies directly on the top of A2 site, while B1 and B2
do not. We notice the bond of B1-B2 is not perpendicular to $sp^2$
orbits, so the coupling between the $\pi$ and $\sigma$ orbits of the
inter-layer B atoms can exist with the strength on the order of
$0.1eV$. Then as illustrated in Fig. 1, the $\pi$ electron (red
circle) at site B1 can hop to $\sigma$ orbits (green circle) at site
B2 and further hops to the B2's $\pi$ orbit with the opposite spin
through the intra-atom ISOC. This hopping process contributes to the
largest ISOC between the $\pi$ electrons in bilayer graphene, with
its strength being on the order of $\Delta$ (not $\Delta^2$), so
this ISOC should be much large than ISOC in a monolayer graphene.

In order to derive the ISOC between $\pi$ electrons in bilayer
graphene, we need to consider one s orbit and three p orbits for
each site, together with two spin indexes, two sub-lattice indexes
and two layer indexes, totally there are 32 orbits in every cell.
Then, the tight-binding Hamiltonian reads:
\begin{equation}
H=\sum\limits_{{\bf i},{\bf j}} \hat{\Psi}^{\dagger}_{\bf
i}\begin{bmatrix}
H_{\sigma, {\bf ij}} &\mathcal{J}_{\bf ij}\\
\mathcal{J}^\dagger_{\bf ij} &H_{\pi, {\bf ij}}
\end{bmatrix} \hat{\Psi}_{\bf j},
\end{equation}
where $\hat{\Psi}_{\bf j}$ ($\hat{\Psi}^{\dagger}_{\bf i}$) is the
annihilation (creation) operators of an electron in the cell ${\bf
j}$ (${\bf i}$) with its dimension $1\times 32$ ($32\times 1$).
$H_{\pi}$, $H_{\sigma}$, and $\mathcal{J}$ are the Hamiltonian of
$\pi$ electrons, $\sigma$ electrons, and the coupling between $\pi$
and $\sigma$ electrons. We adapt the parameters of SWM
model,\cite{ref15,ref16,ref17} and the $\pi$ electron Hamiltonian
$H_{\pi}$ includes the intra-layer N.N. hopping coupling $\gamma_0$,
the interlayer hopping coupling $\gamma_1$ of A1-A2 and that
$\gamma_3$ of B1-B2 (see Fig.1). The value of $\gamma_0=3.12
eV$,$\gamma_1=0.38 eV$,$\gamma_3= 0.29 eV$ is adopted from
Ref.\cite{ref15,ref16,ref17}. The on-site energy of $\pi$ electrons
is set at zero as an energy zero-point. For the $\sigma$ electrons,
we choose the atom orbits $2p_x$, $2p_y$, and $2s$ as the base
vectors. The on-site energies of Hamiltonian $H_{\sigma}$ are 0 for
$2p_x$ and $2p_y$ orbits and $\epsilon_{sp}$
($\epsilon_{sp}=-8.87eV$) for $2s$ orbit. There is no coupling
between $\sigma$ electrons in same atoms because of the orthogonal
base vectors. But the hopping couplings between intra-layer N.N. A-B
atoms are very strong. We first consider a special case that the
bond of A-B atoms is parallel to x axis, thus the hopping
Hamiltonian $H_{\sigma h}$ reads:
\begin{eqnarray}
 H_{\sigma h,  A B} &= & \hat{\Psi}^{\dagger}_{\sigma A}
\begin{bmatrix}
V_{pp \sigma}& 0 &V_{sp}\\
0 &V_{pp\pi}& 0\\
V_{sp}& 0 &V_{ss} \end{bmatrix} \hat{\Psi}_{\sigma B}  +h.c. \nonumber\\
 &\equiv &
\hat{\Psi}^{\dagger}_{\sigma A} V_{hop}
 \hat{\Psi}^{\dagger}_{\sigma B} +h.c.,
\end{eqnarray}
where $\hat{\Psi}_{\sigma A/B} = (\hat{a}_{2px, A/B},
\hat{a}_{2py,A/B}, \hat{a}_{2s,A/B})^T$ is the annihilation
operators of electron in the $2p_x$, $2p_y$, and $2s$ orbits of the
A/B atom. In Eq.(2) the B atom must be the N.N. of A. The hopping
elements $V_{pp\sigma} = 5.04eV$, $V_{pp\pi}=\gamma_0 =3.12 eV$,
$V_{sp} = 5.58 eV$, and $V_{ss} = 6.77 eV$.\cite{ref19} Second, if
the bond of A-B atoms is at an angle $\theta$ to the x axis, by
taking a rotation in xy plane, the hopping Hamiltonian $H_{\sigma
h}$ is obtained as: $H_{\sigma h} = \hat{\Psi}^{\dagger}_{\sigma A}
R^{-1}(\theta) V_{hop} R(\theta) \hat{\Psi}^{\dagger}_{\sigma B}
+h.c.$, with
\begin{align}
R(\theta)=
\begin{bmatrix}
\cos \theta&-\sin \theta& 0\\
\sin \theta&\cos    \theta& 0\\
 0& 0&1
\end{bmatrix}.
\end{align}
The hopping between $\sigma$ orbits of inter-layer is neglected
because that the $\sigma$ electron is mainly within each layer.

The $\pi$-$\sigma$ hopping Hamiltonian $\mathcal{J}$ is from both
ISOC at each atom and the inter-layer $\pi$-$\sigma$ coupling. The
intra-layer $\pi$-$\sigma$ coupling for the N.N. A-B atoms is zero
as mentioned above. The inter-layer $\pi$-$\sigma$ coupling in the
aligning A1 and A2 atoms is also zero since the A1's $\pi$ orbit
points to the center of the A2 atom. But the inter-layer
$\pi$-$\sigma$ coupling between the B1 and B2 atoms exists because
that the bond $B_1$-$B_2$ is not perpendicular to the $2p_x$ and
$2p_y$ orbits. Such kind of $\pi$-$\sigma$ coupling is a crucial
characteristic in bilayer graphene and does not exist in a monolayer
graphene. The Hamiltonian of this $\pi$-$\sigma$ coupling is:
\begin{equation}
 \mathcal{J}_{\pi\sigma } =  \hat{a}^{\dagger}_{pz, B1}
\begin{bmatrix} \gamma'_3 & 0 \end{bmatrix}
\begin{bmatrix} \cos\phi & -\sin\phi\\ \sin\phi &\cos\phi \end{bmatrix}
\begin{bmatrix}\hat{a}_{px,B2}\\ \hat{a}_{py,B2} \end{bmatrix} +h.c.,
\end{equation}
where $\gamma'_3 =\gamma_3 *a/d \approx 0.1eV$ with the lattice
constant $a$ and the distance $d$ between the layers. $\phi$ is the
angle of the projection of the B1-B2 bond in $xy$ plane and the $x$
axis. For every B1 atom there are three N.N. B2 atoms, and their
$\pi$-$\sigma$ couplings are similar except for the different
$\phi$. Similarly, the B2's $\pi$ orbit also has the coupling with
the $\sigma$ orbits of the B1 atoms. Finally, in the 2s and 2p base
vectors, the SOC $\Delta \vec{L}\cdot \vec{\sigma}$ at each atom
site can be easily rewritten as:
\begin{equation}
 \mathcal{J}_{SOC} = \frac{\Delta}{2} \hat{a}^{\dagger}_{pz \alpha}
\begin{bmatrix} \alpha  & i \end{bmatrix}
\begin{bmatrix}\hat{a}_{px\bar{\alpha}}\\ \hat{a}_{py\bar{\alpha}} \end{bmatrix}
 +h.c.,
\end{equation}
where $\alpha=\uparrow,\downarrow$ is the spin index,
$\bar{\alpha}=\downarrow$ ($\uparrow$) for $\alpha=\uparrow$
($\downarrow$), and $i$ is the imaginary unit. Here the SOC is the
spin-flip hopping, but all other hoppings mentioned above are
between the same spin orbits.

Because that the infinity bilayer graphene is a periodic system, we
can rewrite the Hamiltonian in Eq.(1) in the momentum space:
\begin{eqnarray}
H &= &\int\int (dk_x dk_y/4\pi^2) \hat{\Psi}^{\dagger}_{\bf
k}\begin{bmatrix}
H_{\sigma}({\bf k}) &\mathcal{J}({\bf k})\\
\mathcal{J}^\dagger({\bf k}) &H_{\pi}({\bf k})
\end{bmatrix} \hat{\Psi}_{\bf k}\nonumber\\
 &\equiv & \int\int (dk_x dk_y/4\pi^2)
\hat{\Psi}^{\dagger}_{\bf k} H({\bf k}) \hat{\Psi}_{\bf k},
\end{eqnarray}
where the momentum ${\bf k}=(k_x, k_y)$ is in the first Brillouin
Zone.

Next we focus on how to derive the low energy effective Hamiltonian.
First the high-energy $\sigma$ electrons need to be eliminated. For this
purpose, we perform a canonical transformation:
$\widetilde{\mathcal{H}}({\bf k})=e^{-S}H({\bf
k})e^{S}$,\cite{ref18,ref12} where $S=\begin{pmatrix}
0&M\\-M^\dagger& 0
\end{pmatrix}$ is anti-hermitian matrix. After the canonical
transformation, we demand that the $\sigma$ and $\pi$ electrons are
decoupled in the Hamiltonian $\widetilde{\mathcal{H}}({\bf k})$. By
using the condition of the decoupling and keeping the matrix
elements in $\widetilde{\mathcal{H}}_{\pi}$ up to order
$\mathcal{J}^2$, the matrix $M$ can be obtained:
\begin{equation}
 M=-H_\sigma^{-1}\mathcal{J} - H_\sigma^{-2}\mathcal{J} H_{\pi} -\cdots
\end{equation}
While around Dirac points for the low-energy electron, it is safe to
take the approximation $M \approx - H_\sigma^{-1} \mathcal{J}$,
because of large energy gap of several eV between $\sigma$ and $\pi$
electrons. To substitute this $M$ into the above equation of the
canonical transformation, the effective Hamiltonian of $\pi$
electrons with ISOC reads:
\begin{equation}
\widetilde{H_\pi}({\bf k})=H_\pi-2\mathcal{J}^{\dagger}
H_\sigma^{-1}\mathcal{J}
\end{equation}
The magnitude of the ISOC term is roughly estimated as
$\Delta\gamma'_3/\epsilon_{\sigma\pi}$, in which
$\epsilon_{\sigma\pi}$ is the energy difference between $\sigma$ and
$\pi$ bonds at Dirac points.

\begin{figure}
\includegraphics[width=8cm,totalheight=6cm]{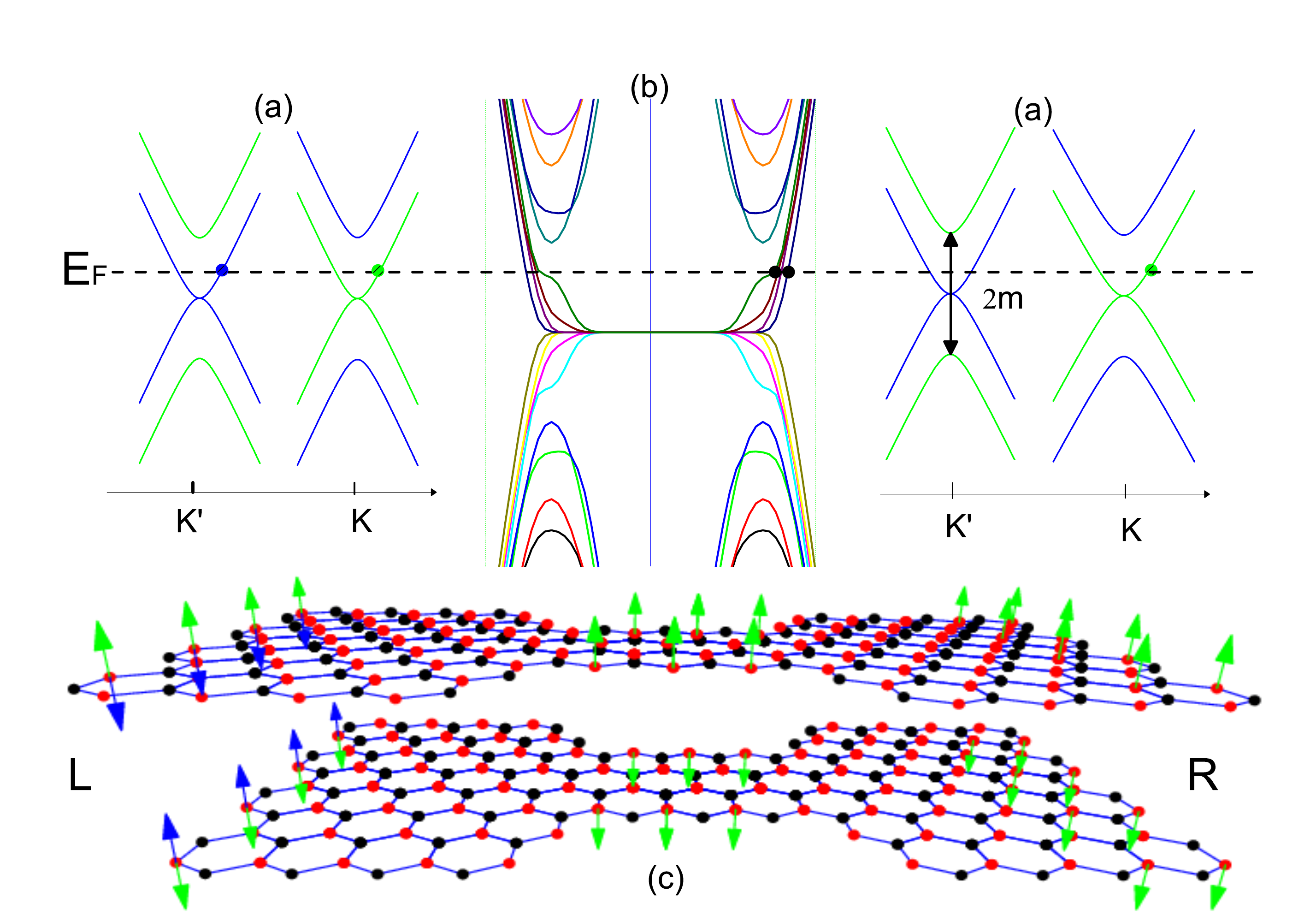}
\caption{(Color online) (a) and (b) are the dispersion relations for
infinite bilayer graphene (set $k_y=0$) and zigzag bilayer graphene
nanoribbon with width $W=120\sqrt{3}a$. (c) is the schematic diagram
for a hybrid device consisting of two large bilayer graphene
connected by a narrow zigzag bilayer graphene ribbon. The green
(blue) arrows in (c) represent the spin polarized direction for the
green (blue) state in (a).}
\end{figure}

In bilayer graphene, the energy bands of $\pi$ electron are fourfold
except for the spin freedom of degree. Two energy bands touch at
Dirac points dominate the low-energy properties, while the other two
bands have an energy gap of about 2$\gamma_1$ due to the strong
interlayer coupling between $\pi$ electrons on site A1 and A2.
Because that the gap $2\gamma_1$ is about $0.7eV$, we need to carry
out the canonical transformation procedure again to eliminate the
high energy $\pi$ bands for sites $A_1$ and $A_2$. Finally, we get
the low-energy effective Hamiltonians  $H_K$ and $H_{K'}$ around the
Dirac points $K$ and $K'$:
\begin{eqnarray}
H_{K}({\bf k}) &= & \left(
\begin{array}{cc}
0 & Z    \\
Z^* & 0 \\
\end{array}
\right) \otimes{\bf I}_{2} +\left(
\begin{array}{cccc}
0 & 0 & 0 & m \\
0 & 0 & 0 & 0 \\
0 & 0 & 0 & 0\\
m & 0 & 0 & 0 \\
\end{array}
\right) \\
H_{K'}({\bf k}) &= & \left(
\begin{array}{cc}
0 & Z^*  \\
Z & 0 \\
\end{array}
\right) \otimes{\bf I}_{2} +\left(
\begin{array}{cccc}
0 & 0 & 0 & 0 \\
0 & 0 & m & 0 \\
0 & m & 0 & 0\\
0 & 0 & 0 & 0 \\
\end{array}
\right)
\end{eqnarray}
where $Z=\{\frac{v_F^2k_+^2}{\gamma_1}+3\gamma_3ak_-\}$, $v_F
=3\gamma_0 a/2$ is the Fermi velocity of $\pi$ electrons, and $k_+ =
k_x + ik_y,k_-=k_x-ik_y$. In the Hamiltonians $H_K$ or $H_{K'}$,
the momentum ${\bf k}$ is measured from $K$ or $K'$, respectively,
and the basis orbits are ($p_{z\uparrow}B1$, $p_{z\downarrow}B1$,
$p_{z\uparrow}B2$, $p_{z\downarrow}B2$). The first term in Eqs.(9)
or (10) comes from the well known low-energy Hamiltonian of $\pi$
electrons on sites $B_1$ and $B_2$ of bilayer
graphene.\cite{ref3,ref15} The second term is the ISOC between the
$\pi$ electrons, which is a central result of this work. This ISOC
origins from the interaction of both the intra-atom ISOC and
interlayer hopping that couples the $\pi$-$\sigma$ orbits as shown in
Fig.1. Its strength $m$ is about
$\Delta\gamma'_3/\epsilon_{\sigma\pi}$, which is on the order of
$\Delta$. From our calculation, the strength $m =0.46meV$ and it is
about 100 times larger than the ISOC in a monolayer graphene.

Eqs (9) and (10) lead to energy bands $E(\vec{k})=
\pm1/2(|m|\pm\sqrt{m^2+4|Z|^2})$. Fig.2a depicts the energy
dispersion at fixed $k_y=0$. Due to the ISOC, the spin degeneracy is
removed. Here two bands have an energy gap $2m$ and the two
low-energy bands still touch at Dirac points. Around the Dirac point
$K$, the low-energy bands are spin polarized. The spins in the upper
layer are mainly upward while they are downward in the lower layer,
as shown in Fig.2c. Moreover, this spin polarization is almost
independent to the momentum $|{\bf k}|$. To compare with the quantum
spin Hall effect, in which the spin polarization is opposite on the
opposite edges, now the opposite spin polarizations exhibit on the
opposite surfaces. Near the Dirac point $K'$, the spin polarization
in the low-energy excitation mode is in contrary to that of $K$
point since they are the time-reversal Krammer pair. Given a
time-reversal invariant impurity, the inter-valley scattering
processes between these two Krammer's pair are
suppressed.\cite{ref20,ref21} Thus, the spin relaxation time is
long.

After obtaining the ISOC term in k-space, its tight-binding form in
real space can be derived straightforwardly. Here we keep the $\pi$
orbits of the A1 and A2 atoms, and the tight-binding Hamiltonian is:
\begin{eqnarray}
 H&
   =& - \sum_{{\bf i},\alpha,\overrightarrow{\delta_i}} \gamma_0 [a_{{\bf i}\alpha}^{\dagger}
 b_{({\bf i}+\overrightarrow{\delta_i}) \alpha} + c_{{\bf i}\alpha}^{\dagger} d_{({\bf i}-\overrightarrow{\delta_i}) \alpha}]
 - \sum_{{\bf i}, \alpha} \gamma_1 a_{{\bf i}\alpha}^{\dagger}c_{{\bf i}\alpha} \nonumber\\
 &-&
 \sum_{{\bf i},\alpha,\overrightarrow{\delta_i}} \gamma_3 b_{{\bf i}\alpha}^{\dagger}d_{({\bf i}+\overrightarrow{\delta_i}) \alpha}
+ \sum_{{\bf i},\alpha,\overrightarrow{\delta_i}}
t_{\vec{\delta_i}\alpha} b_{{\bf i}\alpha}^{\dagger}d_{({\bf
i}+\vec{\delta_i})\overline{\alpha}}+h.c.
\end{eqnarray}
where $a_{{\bf i}\alpha}(a_{{\bf i}\alpha}^\dagger)$, $b_{{\bf
i}\alpha}(b_{{\bf i}\alpha}^\dagger)$, $c_{{\bf i}\alpha}(c_{{\bf
i}\alpha}^\dagger)$, and $d_{{\bf i}\alpha}(d_{{\bf
i}\alpha}^\dagger)$ denotes the annihilation(creation) operator of
the $\pi$ electron at the A1, B1, A2, and B2 atoms with spin
$\alpha$ in the cell ${\bf i}$.
${\overrightarrow{\delta}_{i=1,2,3}}$ are the three N.N. vectors. In
equation (11), the first, second, and third terms express the
intra-layer and inter-layer $\pi$-orbit coupling without spin flip.
The last term leads to the spin flip hopping between the B1's and
B2's $\pi$ orbits due to the ISOC, with
$t_{\vec{\delta}_{i\uparrow}}=t^*_{\vec{\delta}_{i\downarrow}}=\frac{m}{3}exp(i\vec{K}\cdot\vec{\delta}_i)$.
We note that this Hamiltonian maintains $C_3$ symmetry as well as
the time reversal symmetry. By using this tight-binding model,
electronic properties of finite bilayer graphene can be calculated.

Until now, we exhibited that the low-energy modes in bilayer
graphene are spin polarized due to the ISOC. But in the equilibrium
both modes around $K$ and $K'$ are occupied and the total spins are
unpolarized. For the purpose of selecting a single excitation mode,
we need to choose a valley. We design a hybrid system, the so called
`valley filter'\cite{ref22} consisting of two wide bilayer graphenes
connected by a narrow zigzag bilayer graphene nanoribbon (see the
sketch map in Fig.2c), to select a single excitation mode. Earlier
works\cite{ref23,ref22} on monolayer graphene ribbon with zigzag
edges have shown that there exists a propagating chiral edge state,
and the `chirality' means that the sign of group velocity determines
the valley. The situation is the same for the zigzag bilayer
graphene ribbon, as demonstrated from its dispersion relation in
Fig.2b. Let us set the Fermi energy as shown in Fig.2a and 2b and
apply a positive voltage to the device. There will be two kinds of
excitation modes (blue and green arrows in Fig.2c) in the left
region. However, the edge states of the central zigzag ribbon can
only support the one around K point, thus only one excitation mode
(green arrows in Fig.2c) transport to the right region after the
selection, which is spin polarized. Protected by time reversal
symmetry and the large separation between $K$ and $K'$, the spin
polarization in the right region is difficult to be scattered and
can last for a long time. Meanwhile, if we change the direction of
current flow, the other kind of excitation mode can present in the
left region.

In summary, by considering the hopping coupling between interlayer
$\pi$-$\sigma$ orbits and the ISOC in carbon atom, we find a large
ISOC between $\pi$ electrons in bilayer graphene, with its strength
being around $0.46meV$, about 100 times larger than that in the
monolayer graphene. The Hamiltonians of this ISOC in both momentum
and real space are derived. Due to the ISOC, the low-energy electrons
around the Dirac point exhibit a special spin polarization, in which
the electron spins in upper and lower layers are polarized in
opposite directions. In addition, we propose a hybrid system to
select a certain spin polarizing mode by an electric manipulation.

{\bf Acknowledgments:} This work was financially supported by
NSF-China under Grants Nos. 10734110, 10821403, and 10974236,
China-973 program and US-DOE under Grants No. DE-FG02- 04ER46124.

\end{document}